\documentclass[11pt,a4paper]{article}
\pdfoutput=1

\usepackage{jheppub}

\usepackage{amscd,amsmath,amssymb,amsfonts,xspace,mathrsfs,amsthm}
\usepackage{xcolor}

\usepackage{bbold}
\usepackage{latexsym}
\usepackage{graphicx}
\usepackage{tikz-cd}

\usepackage[english]{babel}

\usepackage{graphicx, subfig, float}

\numberwithin{equation}{section}

\newcommand{\refb}[1]{(\ref{#1})}

\newcommand{\sgn}{\mathop{\mathrm{sgn}}}
\newcommand{\Pf}{\mathop{\mathrm{Pf}}}

\title{A simple model for Hawking radiation}

\author{Swapnamay Mondal$^{1,2}$
\\
$^1$ {\it School of Mathematics, Trinity College, Dublin 2, Ireland}
\\
$^2$ {\it Hamilton Mathematical Institute, Trinity College, Dublin 2, Ireland}
\\
$^3$ {\it International Centre for Theoretical Sciences, \\
Tata Institute of Fundamental Research, Shivakote, \\
Hesaraghatta, Bangalore 560089, India}
\vspace*{2mm}\\
{\tt e-mail: swapno@maths.tcd.ie, swapnamay.mondal@icts.res.in, }
\vspace*{-3mm}
}

\abstract{We consider $n$ free Majorana fermions probing a SYK system comprising of $N$ Majorana fermions. We solve the full system in deep infrared and in large $N$ (as well as large $n$) limit. The essential physics of the SYK system is not affected by the probe Majoranas, except addition of another tower of primaries.
The SYK system is seen to induce maximal chaos as well as the whole spectrum of primaries, on to the probe system. The renormalization of soft mode action is computed. We comment on features in common with Hawking radiation.
}

\begin{document}

\maketitle

%%%%%%%%%%%%%%%%%%%%%%%%%%%%%%%%%%%%%%%%%%%
%%%%%%%%%%%%%%%%%%%%%%%%%%%%%%%%%%%%%%%%%%%
\section{Introduction} % \label{s1}
In \cite{Hawking2}, Hawking looked at a free massless scalar field interacting with a black hole in a quantum mechanical manner. It was found that starting with the vacuum state at past null infinity, the field evolves into a thermal state at future null infinity. Since this happens due to interaction with the black hole, in this problem the black hole serves as a heat bath and the quantum field as a probe. These findings proved beyond doubt that quantum mechanically black holes have temperature, something inconceivable for classical black holes. \cite{Hawking2} along with earlier works \cite{Hawking1,Bekenstein1,BCH,Bekenstein2} laid foundations of black hole thermodynamics. Rest is history, which we shall not delve into.

In this paper, we ask a simple minded question: can we develop a simple model that captures some key features of Hawking radiation? The first ingredient of such a model would be a model for black hole itself. Then one needs ``probe''. The only thing of importance about the ``probe" is its interaction with the black hole. For a generic large bath and small probe, one expects the bath to induce its temperature on the probe. When the bath is a black hole, one expects more. To get rid of information paradox \cite{info1,info2}, it is necessary for the probe to carry away information from the black hole. Thus, in a toy model of Hawking radiation, we expect the bath to induce more than just temperature on the probe. Another key feature of black holes is that they saturate the chaos bound \cite{Maldacena:2015waa}, thus it is not unreasonable to expect the black hole to inflict high degree of chaos onto the probe.  

In this work, we present a toy model, meeting above expectations. The first ingredient is a model for black holes, which we take to be the Sachdev-Ye-Kitaev model (henceforth abbreviated as SYK model). This model was proposed by Kitaev  \cite{K2} and by now have been studied extensively \cite{Maldacena:2016hyu, Jensen:2016pah, Engelsoy:2016xyb, Maldacena:2016upp, Garcia-Garcia:2017pzl, Sonner:2017hxc, Fu:2016vas, Hunter-Jones:2017raw, Berkooz:2016cvq, Turiaci:2017zwd, Gross:2016kjj, Polchinski:2016xgd, Gross:2017hcz, Gross:2017aos, Das:2017pif, Bagrets:2017pwq, Larsen:2018iou, Castro:2018ffi}. A drawback of SYK model is that it is not fully quantum mechanical. This led Witten \cite{Witten:2016iux} to point out that certain fermionic tensor models \cite{Gurau:2010ba, Gurau:2011aq} have the same large N physics as SYK model. Such tensor models have also been explored in various directions \cite{Klebanov:2016xxf, Peng:2016mxj, Krishnan:2016bvg, Choudhury:2017tax, Bulycheva:2017ilt, Peng:2017kro, Yoon:2017nig}. 

Previously, a tensor model for a probe to a black hole was proposed in \cite{Halmagyi:2017leq} (also see \cite{Iizuka:2008hg,Iizuka:2008eb, Michel:2016kwn, deBoer:2017xdk}). However original SYK model has a technical advantage over its tensorial counterparts that it is easier to re-express the theory in terms of bilocal fields\footnote{Recently this has been achieved for tensor models as well, using 2PI effective action \cite{Benedetti:2018goh}.}.
This motivates us to take SYK model as the model for black hole in this work. 

Two key features of the SYK model are:
\begin{enumerate}
\item
In deep infrared and large N limit, where SYK model is solvable, it develops an emergent reparameterization symmetry, which is then spontaneously as well as explicitly broken. This broken symmetry is identified with near horizon symmetry of near extremal black holes. 
\item
In the same regime, the model develops maximal chaos \cite{Maldacena:2015waa}, a non trivial feature known to hold for black holes. 
\end{enumerate}
The second ingredient of our model is the ``probe". We take this to be a bunch of free Majorana fermions. The probe keeps both of the above mentioned features of SYK model intact, thus justifying the name ``probe". However a probe to a black hole is supposed to exhibit more exquisite features, as discussed earlier. We find that certain data about the SYK model, namely the spectrum of primary operators, indeed gets imprinted on the probe. Moreover the SYK system inflicts maximal chaos on to the probe.
Due to these non trivial features, we propose this model as a toy model for Hawking radiation.

The paper is organised as follows: 
In section \ref{s2}, we set out by briefly reviewing the SYK model, which we shall take to be the model for black holes. In section \ref{s3} we introduce a model for Hawking radiation and perform a detailed analysis. In \ref{s3.1} we rewrite the theory in terms of bilocal fields. The saddle point equations exhibit reparameterization symmetry, which is spontaneously broken down to $SL(2, \mathbb{R})$ by the saddle point solutions. In \ref{s3.2} we compute one loop effective action around the saddle and derive four point functions therefrom. 
Regulation of otherwise divergent four point function leads to explicit breaking of reparameterization symmetry. Nevertheless the four point functions have conformal parts.
In \ref{s3.3}, the spectrum of primary operators is extracted from these parts. The probe system is seen to develop an identical spectrum of primaries as the ``black hole". This is interpreted as the probe ``copying" some information about the black hole. In \ref{ssoft} we compute the action for soft mode, which turns out to be independent of the probe-black hole coupling. We extract dominant contributions to four point functions, emanating from the soft mode. In \ref{schaos} we determine the Lyapunov exponent for all the four point functions to be maximal. This is interpreted as the black hole inflicting maximal chaos onto the probe. In \ref{ssize} we draw reader's attention to the fact that the saddle point analysis continues to hold when the ``probe" is a larger system than the black hole. Thus an arbitrarily small (in relative terms) system, i.e. the black hole, is capable of inflicting chaos onto an arbitrarily large system, i.e. the ``probe". Lastly, in section \ref{scomment} we make final comments and discuss future directions. 
%%%%%%%%%%%%%%%%%%%%%%%%%%%%%%%%%%%%%%%%%%%%%%%%%%%%%%%%%%%%%%%%%%%%%%%%%%%%%%%%%%%%%%%
\section{A toy model for black holes}  \label{s2}
%%%%%%%%%%%%%%%%%%%%%%%%%%%%%%%%%%%%%%%%%%%%%%%%%%%%%%%%%%%%%%%%%%%%%%%%%%%%%%%%%%%%%%%
We take SYK model \cite{K2} to be the model for black holes and start out by giving a lightening review of this model. 
The SYK model contains $N$ Majorana fermions $\psi_i,~i=1,\dots,N$, where $N$ is a large number. The Hamiltonian is taken to be 
\begin{align}
H_{SYK} &= i^{q/2} \sum_{1 \leq j_1 < \dots < j_q \leq N } j_{j_1 \dots j_q} \psi_{i_1} \dots \psi_{i_q} \, . \label{SYKham}
\end{align}
Here $ j_{j_1 \dots j_q} $ are random couplings to be averaged over an ensemble, specified by
\begin{align}
\langle j_{i_1 \dots i_q} \rangle &= 0, ~ \langle j^2_{i_1 \dots i_q} \rangle = \frac{J_0^2 (q-1)! }{N^q} \, . 
\end{align}
This model turns out to be solvable at large $N$ limit, in deep infrared region. There are two equivalent approaches to solve the problem. The first one uses diagrammatic techniques and the second one involves integrating original fermions out in favour of certain bilocal fields. We shall follow the second approach in section \ref{s3}. But in this section use diagrammatic techniques.

\begin{figure}[H]
	\begin{center}
		\includegraphics[scale=0.7]{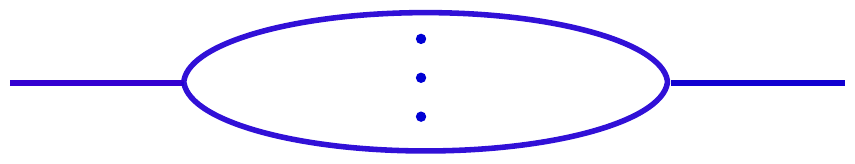}
		\caption{Leading contributions to the propagator in large $N$ limit.}	\label{psiprop}
	\end{center}
\end{figure}

In the limit mentioned above, leading diagrams are ``melonic" and can be summed over without much difficulty.  One starts by noting that the following contribution (see figure \ref{psiprop}) to the propagator $G^\psi(t) := \frac{1}{N} \sum_i \langle T \psi_i(t) \psi_i(0) \rangle$ is of order $N^0$. The blue lines in figure \ref{psiprop} represent a $\psi$ field propagator. In the remaining part of the paper, blue lines will continue to represent $\psi$ field. Any new field would be represented by a different color.
This means if one replaces any internal line by figure \ref{psiprop} itself, one has a new order $N^0$ contribution. This simple pattern allows one to sum up all such diagrams. Upon checking that such diagrams indeed cover all leading contributions in $1/N$ expansion, one can arrive at the Schwinger-Dyson equation
\begin{align}
J_0^2\int dt~G^\psi(t_1,t) \left( G^\psi(t,t_2) \right)^{q-1} &= -\delta(t_1 - t_2) \, . \label{SDpsi}
\end{align}
This equation is invariant under the conformal\footnote{In 1 dimension any reparameterization is a conformal transformation.} transformations
\begin{align}
G^\psi(t_1,t_2) \rightarrow \left| \frac{df(t_1)}{dt_1} \frac{df(t_2)}{dt_2} \right|^{1/q} G^\psi(f(t_1), f(t_2)) \, . \label{conf}
\end{align}
However the solution to this equation 
\begin{align}
G_c(t) &= \frac{b}{|t|^{2/q}} \sgn(t),~~~~\text{where,}~~J_0^2 b^q \pi  = \left( \frac{1}{2} - \frac{1}{q} \right) \tan \frac{\pi}{q} \, , \label{SYKpropa}
\end{align}
spontaneously breaks this symmetry down to $SL(2,\mathbb{R})$. This pattern of symmetry breaking is analogous to that of $AdS_2$ space, whose asymptotic symmetry group contains all reparameterizations of the boundary circle whereas the $AdS_2$ metric preserves only a $SL(2, \mathbb{R})$ subgroup. Spontaneous breaking of a symmetry also implies existence of Goldstones, which should be particularly important for low energy physics. However there is more to the symmetry breaking in the SYK model, as we shall find shortly.
%%{\color{gray} Given that $AdS_2$ emerges as near horizon geometry of extremal black holes, one might wonder if SYK model can be thought of as a model for extremal black holes. However analysis of four point functions shows that the conformal symmetry is further broken explicitly. This is the pattern of symmetry breaking of nearly $AdS_2$ spaces \cite{Maldacena:2016upp}, which in turn arises as near horizon geometry of near extremal black holes. Apart from resemblance in pattern of symmetry breaking, SYK model shares another non-trivial property of black holes, namely maximal chaos. Altogether SYK model stands some chance of being a good model for near-extremal black holes.} %(or more optimistically an example of so called near-$AdS_2$/near-$CFT_1$ duality).

It is instructive to note that $G_c(t)$ in \refb{SYKpropa} admits a smooth $J_0 \rightarrow \infty$ limit, but not a smooth $J_0 \rightarrow 0$ limit. This implies that deep infra red limit is a strong coupling limit.

\begin{figure}[H]
	\begin{center}
		\includegraphics[scale=0.7]{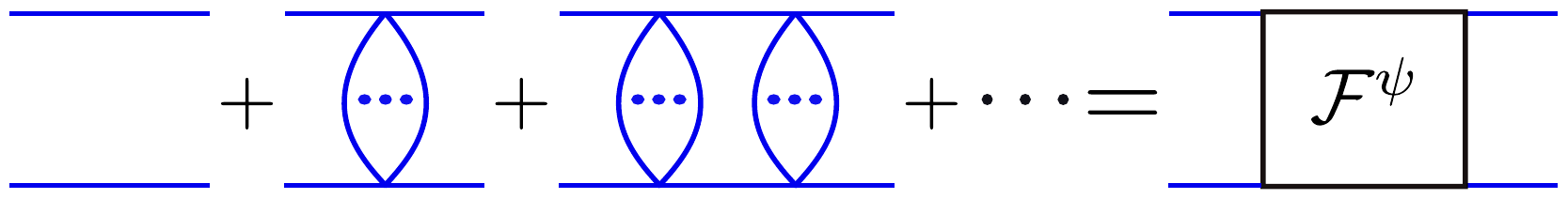}
		\caption{Leading contributions to ``gauge invariant four point function" in large $N$ limit.}	\label{klebladder}
	\end{center}
\end{figure}

Coming to four point functions, ``gauge invariant" four point function has the following structure
\begin{align}
\frac{1}{N^2} \sum_{i,j} \langle \psi_i(t_1)  \psi_i(t_2)  \psi_j(t_3) \psi_j(t_4) \rangle &= G^\psi(t_1,t_2) G^\psi(t_3,t_4) + \frac{1}{N} \mathcal{F}^\psi(t_1,t_2;t_3,t_4) \, , \label{psi4defn}
\end{align} 
where $\mathcal{F}^\psi$ is given by the sum of ladder diagrams shown in figure \ref{klebladder}.

A ladder with $n$ rungs is denoted as $\mathcal{F}^\psi_n$ and can be obtained from $\mathcal{F}^\psi_{n-1}$ by acting with the kernel $K_c$:
\begin{align}
\nonumber
\mathcal{F}^\psi_n(t_1,t_2;t_3,t_4) &= \int dt dt'~K_c(t_1,t_2; t, t') \mathcal{F}^\psi_{n-1}(t, t'; t_3,t_4) \, ,\\
\text{where}~~~K_c(t_1,t_2; t_3,t_4) &= - (q-1) J_0^2 G_c(t_1,t_3) G_c(t_2,t_4) G_c(t_3,t_4)^{q-2} \, .
\label{kleb_frecursion}
\end{align}
The kernel $K_c$ commutes with $SL(2,\mathbb{R})$ generators. Given any generator $J$ of $SL(2,\mathbb{R})$, one has
\begin{align}
(J_1 + J_2) K_c(t_1,t_2;t_3,t_4) &= K_c(t_1,t_2;t_3,t_4) (J_3 + J_4) \, .
\label{kleb_kcommutes}
\end{align}
Here $J_i$ acts at time $t_i$. Using \refb{kleb_frecursion} one can sum up the ladder diagrams to obtain
\begin{align}
\mathcal{F}^\psi &= (1+ K_c + \left( K_c \right)^2 + \dots ) \mathcal{F}^\psi_0 = \frac{1}{1-K_c} \mathcal{F}^\psi_0 \, . \label{kleb_fsum}
\end{align}
$SL(2,\mathbb{R})$ symmetry \refb{kleb_kcommutes} of the kernel $K_c$ and the fact that 
\begin{equation}
\mathcal{F}^\psi_0(t_1,t_2;t_3,t_4) \equiv - G_c(t_1,t_3) G_c(t_2,t_4) + G_c(t_1,t_4) G_c(t_2,t_3)
\end{equation}
preserves the $SL(2,\mathbb{R})$ symmetry, seems to suggest that $\mathcal{F}^\psi$ preserves $SL(2,\mathbb{R})$ symmetry as well. However this is not the case, since $K_c$ happens to have a unit eigenvalue, implying $\mathcal{F}^\psi(t_1,t_2;t_3,t_4)$ diverges in the strict conformal limit. To take care of this divergence, it is necessary to move away slightly from the conformal point. This leads to explicit breaking of conformal symmetry.
Thus the emergent reparameterization symmetry in SYK model is broken both spontaneously as well as explicitly. Explicit breaking entails there are no true Goldstones, but only what may be called pseudo-Goldstones. This particular pattern of symmetry breaking is also exhibited by nearly AdS$_2$ spaces \cite{Maldacena:2016upp}, which in turn arises as near horizon geometry of nearly extremal black holes. This already inspires curiosity about the possibility of SYK model being a model for near extremal black holes. But we shall wait for one final piece: maximal chaos.

Before that, we mention $\mathcal{F}^\psi(t_1,t_2;t_3,t_4)$ takes the following form after regulating the divergence:
\begin{align*}
\mathcal{F}^\psi(t_1,t_2;t_3,t_4) &= \mathcal{F}_{non-conf}^\psi(t_1,t_2;t_3,t_4) + G^\psi(t_1,t_2) G^\psi (t_3,t_4) \mathcal{F}^\psi(\chi) \, ,
\end{align*}
where $\mathcal{F}^\psi_{non-conf}(t_1,t_2;t_3,t_4)$ is the conformal symmetry breaking piece, $G^\psi(t_1,t_2) G^\psi (t_3,t_4) \mathcal{F}^\psi(\chi)$ is the conformal symmetry preserving piece and $\chi := \frac{t_{12}t_{34}}{t_{13} t_{24}}$ is the $SL(2,\mathbb{R})$ invariant cross ratio. $\mathcal{F}^\psi(\chi)$ can be evaluated utilising conformal symmetry \cite{Maldacena:2016hyu}. Of particular interest, is the $\chi \rightarrow 0$ limit. In this limit, one has
\begin{align}
\mathcal{F}^\psi (\chi) &=  \sum_{m=1}^\infty \alpha_0 \frac{(h_m-1/2)}{\pi \tan (\pi h_m/2)} \frac{\Gamma(h_m)^2}{\Gamma(2h_m)} \frac{1}{(k'(h_m))} \chi^{h_m} )
 \,  .\label{cm}
\end{align}
Here $h_m$-s are the roots of the equation $k_c(h)=1$, with
\begin{align}
k_c(h) &= - (q-1) \frac{\Gamma( \frac{3}{2}- \frac{1}{q} ) ~\Gamma(1- \frac{1}{q}  )~\Gamma( \frac{1}{q} + \frac{h}{2}  )~\Gamma( \frac{1}{2} + \frac{1}{q}  - \frac{h}{2} )}{\Gamma( \frac{1}{2} + \frac{1}{q} ) ~\Gamma(\frac{1}{q} )~\Gamma( \frac{3}{2} - \frac{1}{q} -\frac{h}{2} )~\Gamma(1-\frac{1}{q} + \frac{h}{2} )} \, , \label{kch}
\end{align}
and $\alpha_0 := \frac{1}{(q-1)J_0^2 b^q}$ is a numerical constant. \refb{cm} implies a tower of conformal primaries with dimensions $h_m$. These dimensions are roughly equally spaced. 

The piece $\mathcal{F}^\psi_{non-conf}(t_1,t_2;t_3,t_4)$ is of $\mathcal{O}(J_0)$ and since we are working in strong coupling regime, this gives dominant contribution to $\mathcal{F}^\psi(t_1,t_2;t_3,t_4)$.  Following \cite{Maldacena:2016hyu}, we shall refer to the dominant piece as $\mathcal{F}_{big}$. Taking $1/J_0$ corrections to the kernel $K_c$ into account,  $\mathcal{F}_{big}$ is evaluated to be (in Euclidean signature)
\begin{align}
\mathcal{F}_{big}(\tau_1,\tau_2;\tau_3,\tau_4) &= \frac{6 \alpha_0 \mathcal{J} \beta}{\pi^2 \alpha_K} \sum_{|m| \geq 2} \frac{e^{-2i\pi m (y_{12} - y_{34})/\beta}}{m^2 (m^2-1)} f_m (x_{12}) f_m(x_{34}) \, ,\label{fbig}
\end{align}
where $\mathcal{J}= 2^{(1-q)/2} \sqrt{q} J_0$ and $\alpha_K$ is a $q$-dependent numerical constant, which approaches the value $3$ from below as $q\rightarrow \infty$. For more details of $\alpha_K$, see \cite{Maldacena:2016hyu}.
$y_{ij}= \frac{\tau_i + \tau_j}{2}, x_{ij}= \frac{\tau_i - \tau_j}{2}, f_n(x)=- n \cos \frac{2\pi nx}{\beta} +  \frac{\sin \frac{2\pi nx}{\beta}}{\tan \frac{\pi \tau}{\beta}}$.

Now we come to analysis of chaos. Chaos is diagnosed by Lyapunov exponent, the rate at which out of time order correlators (OTOC) grow at early times. We shall look at the following out of time order four point function,
\begin{align*}
F(t) &:= \frac{1}{N^2} \sum_{j,k} \langle \psi_j(t+ \frac{3i\beta}{4}) \psi_k(\frac{i\beta}{2}) \psi_j(t+ \frac{i\beta}{4}) \psi_j(0) \rangle_L \, ,
\end{align*}
the subscript ``L" denotes the Lorentzian signature. This can be obtained by first evaluating the Euclidean correlator 
\begin{align}
F(\tau) &:= \frac{1}{N^2} \sum_{j,k} \langle \psi_j(\tau - \frac{3\beta}{4}) \psi_k(-\frac{\beta}{2}) \psi_j(\tau - \frac{\beta}{4}) \psi_j(0) \rangle_E \, , \label{2bev}
\end{align}
and then analytically continuing $\tau \rightarrow it$. We shall omit the subscript ``E", since we will be working in Euclidean signature unless mentioned otherwise. Using \refb{fbig}, the dominant contribution to this is 
\begin{align}
\nonumber
F_{big}(\tau) &= \frac{12 \alpha_0 \mathcal{J} \beta}{ \pi^2 \alpha_K } G_c(\beta/2) G_c(\beta/2) \sum_{n \geq 2, \atop n \in 2 \mathbb{Z}} \frac{(-1)^{n/2} \cos \frac{2\pi n \tau}{\beta}}{(n^2-1)} \\
&=  \frac{12 \alpha_0 \mathcal{J} \beta}{ \pi^2 \alpha_K } G_c(\beta/2) G_c(\beta/2) \left[ \frac{1}{4 i} \lim_{\epsilon \rightarrow 0} \int_{\epsilon - i \infty}^{\epsilon + i \infty} \frac{\cos \frac{2\pi \tau \omega}{\beta}}{\sin \frac{\pi \omega}{2} (\omega^2 -1)}  - \frac{\pi}{4} \cos \frac{2\pi \tau}{\beta} \right] 
\end{align}
The integral converges for $Re(\tau) < \beta/4$. Now we continue to $\tau \rightarrow it$, which keeps the convergence intact. Note that  $t$ appears in the integrand in an oscillatory fashion, thus large $t$ growth is controlled by the second term, which goes like $\cos \frac{2\pi \tau}{\beta} \rightarrow \cosh \frac{2\pi t}{\beta} \sim e^{\frac{2\pi |t|}{\beta}}$. This gives the Lyapunov exponent to be $2\pi/ \beta$.

%which can be obtained by analytically continuing Euclidean four point functions. $\mathcal{F}^\psi_{big}$, continued appropriately, gives dominant contribution to this. The growth of $\mathcal{F}^\psi_{big}$ dictates the Lyapunov exponent to be maximal \cite{Maldacena:2016hyu}. 

%These features suggest SYK model can be regarded as a good model for near extremal black holes.

%For the rest of the paper, we will suppress the $\psi$ superscript in $G^\psi, K^\psi, \mathcal{F}^\psi$ and so on, i.e. $G(t_1,t_2)$ would denote $G^\psi(t_1,t_2)$ and so on. 
%%%%%%%%%%%%%%%%%%%%%%%%%%%%%%%%%%%%%%%%%%%%%%%%%%%%%%%%%%%%%%%%%%%%%%%%%%%%%%%%%%%%%%%%%%%%%%%%%%%%%%%%%%%%%%%%%%%%%%%%%%%%%%%%%%%%%%%%%%%%%%%%%%%%%%%%%%%%%%%%%%%%%%%%%%%%%%%%%%
\section{A toy model for Hawking radiation} \label{s3}
Now we present a class of models, describing a probe interacting with a black hole. The black hole is described by the Hamiltonian \refb{SYKham} of SYK model and interaction with the probe is represented by a new piece $H_{probe}$. This piece contains original $\psi$ fields, representing degrees of freedoms of the black hole, as well as new fields $\kappa_x,~x=1,\dots,n$, representing the probe degrees of freedom. 
$n$ is a large number, which for now will be thought of as being much smaller than $N$. We would consider physics of this model up to leading order in $1/N$ and $1/n$.

$H_{probe}$ is given by
\begin{align}
H_{probe} &= i^{q/2} \sum_{1 \leq i_1 < \dots < i_{q-p} \leq N; \atop 1 \leq x_1 < \dots < x_p \leq n} j'_{i_1 \dots i_{q-p}; x_1 \dots x_p} \psi_{i_1} \dots \psi_{i_{q-p}} \kappa_{x_1} \dots \kappa_{x_p} \, . \label{probeH}
\end{align}
$2 \leq p < q$ is an even number.
$j'_{i_1 \dots i_{q-p}; x_1 \dots x_p}$ are random couplings and have to be averaged over. Disorder average is specified by 
\begin{align}
\langle  j'_{i_1 \dots i_{q-p}; x_1 \dots x_p}  \rangle &=0 ~,~ \left \langle \left( j'_{i_1 \dots i_{q-p}; x_1 \dots x_p}  \right)^2 \right \rangle = \frac{J_1^2 (q-p)! (p-1)! }{n^{p-1} N^{q-p}}  \, .
\end{align}
$H_{probe}$ is a special case of the generalised SYK models considered in \cite{Gross:2016kjj}. However the generalisation of $H_{probe}$ considered in \cite{Gross:2016kjj} was the full Hamiltonian by itself, whereas in the present case $H_{probe}$ describes a probe to SYK system. 
 
\begin{figure}[H]
	\begin{center}
		\includegraphics[scale=0.7]{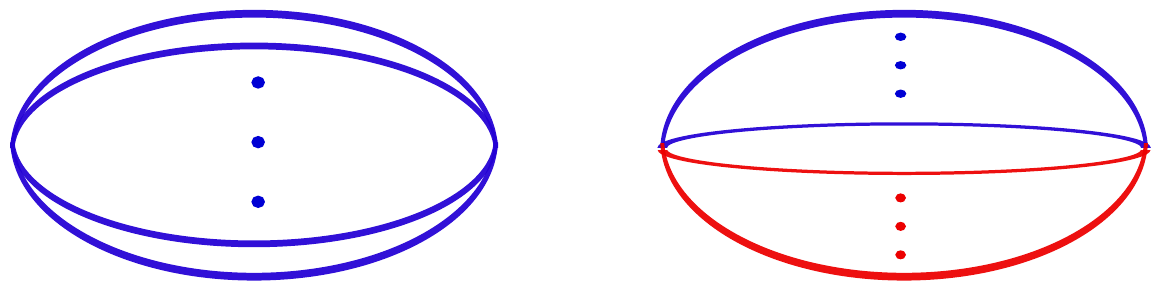}
		\caption{Left: Simplest leading contributions to free energy from \refb{SYKham}. Right: Simplest leading contributions to the free energy from \refb{probeH}.}	\label{vacua}
	\end{center}
\end{figure}

To see that $\kappa_x$ can really be thought of as probes, consider the simplest contributions of $H_{SYK}$ and $H_{probe}$ to the free energy. These are represented by diagrams in figure \ref{vacua}. Blue lines continue to represent $\psi$ propagators and red lines represent $\kappa$ propagators.
It is easy to check that the left diagram is $\mathcal{O}(N)$ and the right one is $\mathcal{O}(n)$. This indicates that introduction of $\kappa$ fields does not alter thermodynamic properties of the system in $n/N <<1, N \rightarrow \infty$ limit. Same holds for correlation functions since many of them are obtained by cutting lines from the vacuum melons. We will see this explicitly in what follows.
%%%%%%%%%%%%%%%%%%%%%%%%%%%%%%%%%%%%%%%%%%%
\subsection{Two point functions} \label{s3.1}
%%%%%%%%%%%%%%%%%%
Performing disorder average over the path integral and then integrating original Majoranas out give an effective action in terms of bilocal fields. This is particularly interesting achievement, because bilocal fields, being gauge invariant, are analogs of Wilson lines in $0+1$ dimension and thus this amounts to rewriting the theory in terms of Wilson lines. Such a rewriting, although physically desirable, has not been achieved in higher dimensions, to the best of our knowledge. Therefore achievability of the same, although in certain limits, is quite significant. 

Upon disorder averaging, one has
\begin{align}
\left \langle e^{-\int d\tau \left( \frac{1}{2} \psi_i \dot{\psi}_i + \frac{1}{2} \kappa_x \dot{\kappa}_x + H \right) }\right \rangle &= \left \langle e^{-\int d\tau \left( \frac{1}{2} \psi_i \dot{\psi}_i + H_{SYK} \right) } \right \rangle \left \langle e^{-\int d\tau \left( \frac{1}{2} \kappa_x \dot{\kappa}_x + H_{probe} \right)} \right \rangle =: e^{-N S^{(1)}_{eff}  -  n  S^{(2)}_{eff}} \, ,
\end{align}
where $\langle~\rangle$ represents disorder average. The first factor can be borrowed from literature and reads, to leading order in $1/N$
\begin{align}
S^{(1)}_{eff}[G_\psi,\Sigma_\psi] &=  -\log \Pf (\partial_\tau - \Sigma_\psi) + \frac{1}{2} \int_0^\beta \int_0^\beta d\tau d\tau' \Big\{ \Sigma_\psi(\tau, \tau') G_\psi(\tau,\tau') - \frac{J_0^2}{q} \left( G_\psi(\tau, \tau')\right)^q \Big\}  \, , \label{s1eff}
\end{align}
where $G_\psi(t_1,t_2) = \frac{1}{N} \sum_i \psi_i(t_1) \psi(t_2)$ (not to be confused with the two point function $G^\psi$) is a bilocal field and $\Sigma_\psi(t_1,t_2)$ is a Lagrange multiplier field. The second factor is given by
\begin{align}
S^{(2)}_{eff}[G_\kappa,\Sigma_\kappa] &=  -\log \Pf (\partial_\tau - \Sigma_\kappa) + \frac{1}{2} \int_0^\beta \int_0^\beta d\tau d\tau' \Big\{ \Sigma_\kappa(\tau, \tau') G_\kappa(\tau,\tau') - \frac{J_1^2}{p} \left( G_\psi(\tau, \tau')\right)^{q-p} \left( G_\kappa(\tau, \tau')\right)^p  \Big\}  \, , \label{s2eff}
\end{align}
where $G_\kappa (t_1,t_2) = \frac{1}{n} \sum_x \kappa_x(t_1) \kappa_x(t_2)$ (not to be confused with the two point function $G^\kappa$) is a bilocal field and $\Sigma_\kappa(t_1,t_2)$ is a Lagrange multiplier field. 
There are two large numbers in the game, $n$ and $N$, with $N>>n$. 

Saddle point equations for $G_\psi$ and $G_\kappa$ are respectively given by 
\begin{align}
\nonumber
%\frac{\partial (-N S^{(1)}_{eff}  -  n  S^{(2)}_{eff})}{\partial G_\psi(\tau, \tau')} &=0  ~\Rightarrow~
%\frac{N}{2} \left( \Sigma_\psi(\tau, \tau') - J_0^2 \left( G_\psi(\tau, \tau')\right)^{q-1} \right) + \frac{n}{2} (\frac{q}{p} -1) \left( G_\psi(\tau, \tau')\right)^{q-p-1} \left( G_\kappa(\tau, \tau')\right)^p  &= 0 \\
 \Sigma_\psi(\tau, \tau') &= J_0^2 \left( G_\psi(\tau, \tau')\right)^{q-1} + \frac{n}{N}  (\frac{q}{p} -1) J_1^2 \left( G_\psi(\tau, \tau')\right)^{q-p-1} \left( G_\kappa(\tau, \tau')\right)^p  \\
%\frac{\partial (-N S^{(1)}_{eff}  -  n  S^{(2)}_{eff})}{\partial G_\kappa(\tau, \tau')} &=0  ~\Rightarrow~ 
\Sigma_\kappa(\tau, \tau') &= J_1^2  \left( G_\psi(\tau, \tau')\right)^{q-p} \left( G_\kappa(\tau, \tau')\right)^{p-1} \, . \label{SDsaddleg}
\end{align}
The similarity of both equations in form motivates one to make the ansatz
\begin{align}
G_\kappa &= \alpha G_\psi \, , \label{ansatz}
\end{align}
for the saddle point values.
Then we have 
\begin{align}
\Sigma_\psi &= \left( J_0^2 + \frac{n}{N} (\frac{q}{p}-1) \alpha^p  J_1^2 \right) G_\psi^{q-1} ~\, , ~
\Sigma_\kappa = J_1^2 \alpha^{p-1}  G_\psi^{q-1} \, . \label{sigmasaddle}
\end{align}
On the other hand, in deep infrared, saddle point equations for $\Sigma_\psi$ and $\Sigma_\kappa$ read respectively 
\begin{align}
\int d\tau~ G_\psi(\tau_1, \tau) \Sigma_\psi(\tau, \tau_2) &= - \delta(\tau_1 - \tau_2) \, ~~\text{and}~~
\int d\tau~ G_\kappa (\tau_1, \tau) \Sigma_\kappa (\tau, \tau_2) = - \delta(\tau_1 - \tau_2) \, . \label{SDsaddlesigma}
\end{align}
Putting \refb{ansatz} and \refb{sigmasaddle} into \refb{SDsaddlesigma} yields two equations for $G_\psi$, which are the same except over all factors. For consistency these factors must be the same. This gives
\begin{align}
%J_1^2 \alpha^p &= J_0^2 -\frac{n}{N} (\frac{q}{p}-1) J_1^2 \alpha^p \, \\
\alpha^p  &= \left( \frac{J_0}{J_1}\right)^2 \left[ 1- \frac{n}{N} \left( \frac{q}{p} -1 \right) \right]^{-1}   \, . \label{alphafix}
\end{align}
Solutions to \refb{SDsaddlesigma} are 
\begin{align}
\nonumber
G^{saddle}_\psi(t) &= \left[ 1 - \epsilon \left( \frac{q}{p} - 1 \right)\right]^{1/q} G_c(t) \\
%\left[ \frac{1}{\pi J^2} \left( \frac{1}{2} - \frac{1}{q} \right) \tan \frac{\pi}{q} \right]^{1/q} \frac{\sgn{} (t)}{|t|^{2/q}} = \left[ \frac{1}{\pi J_0^2} \left[ 1- \frac{n}{N} \left( \frac{q}{p} -1 \right) \right] \left( \frac{1}{2} - \frac{1}{q} \right) \tan \frac{\pi}{q} \right]^{1/q} \frac{\sgn{} (t)}{|t|^{2/q}} \\
G^{saddle}_\kappa(t) &= \left( \frac{J_0}{J_1} \right)^{2/p}  \left[ 1 - \epsilon \left( \frac{q}{p} - 1 \right)\right]^{1/q -1/p} G_c(t)  \, ,\label{saddlesol}
%\left( \frac{J_0}{J_1}\right)^{2/p} \left[ 1- \frac{n}{N} \left( \frac{q}{p} -1 \right) \right]^{-1/p} G^{saddle}_\psi(t)\\
%&= J_1^{-2/p}  \left[ \frac{1}{J_0^2}\left\{ 1- \frac{n}{N} \left( \frac{q}{p} -1 \right) \right\} \right]^{1/q-1/p}  \left[  \left( \frac{1}{2} - \frac{1}{q} \right) \tan \frac{\pi}{q}  \right]^{1/q}  \frac{\sgn{} (t)}{|t|^{2/q}} 
\end{align}
where $\epsilon=n/N$ and $G_c(t)$ is the two point function of the SYK model, given in  \refb{SYKpropa}.

Let us mention couple of noteworthy features. Firstly, $G^{saddle}_\kappa(t)$ admits an expansion around $\frac{J_0}{J_1} \rightarrow 0$ (with $J_0$ held fixed), suggesting that we are in $J_1 >> J_0$ regime.
So the probe couples to the bath more strongly than the bath couples to itself! Usually this is not what one means by probe, nevertheless for the lack of a better name, we shall continue to refer the $\kappa$ system as probe. 

Second curious feature is the $\epsilon$ dependence of the propagators: when expanded around $\epsilon=0$, both two point functions in \refb{saddlesol} contain all powers of $\epsilon$. This implies reproducing \refb{saddlesol} from diagrammatic techniques will require computing infinitely many classes of diagrams. We shall not undertake this potentially never-ending endeavour in this paper, but only reproduce \refb{saddlesol} to first subleading order from diagrammatic techniques in appendix \ref{app}. 
The same feature leads to another just concern: after first few terms in $\epsilon$ expansion, every term would be dominated by some $1/N$ or $1/n$ corrections, i.e. ``quantum effects". Thus is it consistent to keep $\epsilon$ corrections coming from $\left[ 1 - \epsilon \left( \frac{q}{p} - 1 \right)\right]$, but not the comparable/ dominant ones due to ``quantum" effects? We shall take the following point of view regarding this: the key feature of an order by order expansion, is its consistent truncation at any order. Typically mixing up contributions from different orders does not lead to a consistent over all picture. 
In the rest of the paper, we show this is not the case for us though. Nowhere do we truncate the fractional powers of $\left[ 1 - \epsilon \left( \frac{q}{p} - 1 \right)\right]$ and yet we get mathematically consistent and physically sensible answers (e.g. four point functions, action for soft mode).\\
{\color{red}anything more?}
%%%%%%%%%%%%%%%%%%
\subsection{Four point functions} \label{s3.2}
%%%%%%%%%%%%%%%%%%
In order to obtain the four point functions, we expand the effective action around the saddle. To this end let us define
\begin{align}
\nonumber
G_\psi &= G^{saddle}_\psi + \left| G^{saddle}_\psi \right|^{1-q/2} g_\psi \, , \\
\nonumber
G_\kappa &= G^{saddle}_\kappa + \left| G^{saddle}_\kappa \right|^{1-p/2} \left| G^{saddle}_\psi \right|^{\frac{p-q}{2}}  g_\kappa \, , \\
\nonumber
\Sigma_\psi &= \Sigma^{saddle}_\psi + \left| G^{saddle}_\psi \right|^{-1+q/2} \sigma_\psi \, , \\
\Sigma_\kappa &= \Sigma^{saddle}_\kappa + \left| G^{saddle}_\kappa \right|^{-1+p/2} \left| G^{saddle}_\psi \right|^{\frac{q-p}{2}}  \sigma_\kappa \, . \label{saddlexp}
\end{align}
One can check $dG_\psi d\Sigma_\psi = dg_\psi d\sigma_\psi$ and $dG_\kappa d\Sigma_\kappa = dg_\kappa d\sigma_\kappa$. Thus the measure in the path integral over bi-local fields can be replaced by $D\sigma_\psi D\sigma_\kappa Dg_\psi Dg_\kappa$.  

\refb{saddlexp} entails
\begin{align}
\nonumber
\mathcal{F}^\psi (t_1,t_2;t_3,t_4) := N\langle G_\psi (t_1,t_2) G_\psi(t_3,t_4) \rangle_{con} &= N \left| G^{saddle}_\psi(t_1,t_2) G^{saddle}_\psi(t_3,t_4) \right|^{1-q/2} \langle g_\psi(t_1,t_2) g_\psi(t_3,t_4) \rangle \, , \\
\nonumber
\mathcal{F}^\kappa (t_1,t_2;t_3,t_4) := n \langle G_\kappa (t_1,t_2) G_\kappa (t_3,t_4) \rangle_{con} &=  n \left| G^{saddle}_\psi (t_1,t_2) G^{saddle}_\psi (t_3,t_4) \right|^{(p-q)/2} \\
\nonumber
&\left| G^{saddle}_\kappa (t_1,t_2) G^{saddle}_\kappa (t_3,t_4) \right|^{1-p/2} \langle g_\kappa (t_1,t_2) g_\kappa (t_3,t_4) \rangle \, , \\
\nonumber
\mathcal{F}^{\psi \kappa} (t_1,t_2;t_3,t_4) := N \langle G_\psi (t_1,t_2) G_\kappa (t_3,t_4) \rangle_{con} &= N \left| G^{saddle}_\psi (t_1,t_2) \right|^{1-q/2} \left| G^{saddle}_\psi (t_3,t_4) \right|^{(p-q)/2} \\
&  \left| G^{saddle}_\kappa (t_3,t_4) \right|^{1-p/2} \langle g_\psi (t_1,t_2) g_\kappa (t_3,t_4) \rangle \, . \label{Gg}
\end{align}
In order to evaluate these, we set out by expanding the effective actions $S^{(1)}_{eff}$ \refb{s1eff} and $S^{(2)}_{eff}$ \refb{s2eff} around the saddle point to quadratic order in $\sigma_\psi, \sigma_\kappa, g_\psi, g_\kappa$. This yields
\begin{align}
\nonumber
\delta S^{(1)}_{eff} &= -\frac{ 1 - \epsilon \left( \frac{q}{p} -1 \right) }{4J_0^2 (q-1)} \int dt_{1,2,3,4}~ \sigma_\psi (t_1,t_2) \widetilde{K}_c(t_1,t_2;t_3,t_4) \sigma_\psi(t_3,t_4) + \frac{1}{2} \int dt_{1,2}~ g_\psi(t_1,t_2) \sigma_\psi(t_1,t_2) \\
\nonumber
&- \frac{J_0^2 (q-1)}{4} \int dt_{1,2}~ g_\psi^2(t_1,t_2) \, , \\
\nonumber
\delta S^{(2)}_{eff} &= -\frac{1}{4J_1^2 (q-1)} \int dt_{1,2,3,4}~ \sigma_\kappa(t_1,t_2) \widetilde{K}_c (t_1,t_2;t_3,t_4) \sigma_\kappa(t_3,t_4) + \frac{1}{2} \int dt_{1,2}~ g_\kappa(t_1,t_2) \sigma_\kappa(t_1,t_2) \\
\nonumber
&- \frac{J_1^2 (p-1)}{4} \int dt_{1,2}~ g_\kappa^2(t_1,t_2) - \frac{J_1^2 (q-p) (q-p-1) \alpha^p}{4p} \int dt_{1,2}~g_\psi^2(t_1,t_2) \\
&- \frac{J_1^2 (q-p) \alpha^{p/2}}{2} \int dt_{1,2}~g_\psi(t_1,t_2) g_\kappa(t_1,t_2) \, , 
\end{align}
where $\widetilde{K}_c$ is the symmetric kernel 
\begin{align}
\widetilde{K}_c(t_1,t_2;t_3,t_4) =- J_0^2 (q-1) |G_c(t_1,t_2)|^{q/2-1} G_c (t_1,t_3) G_c(t_2,t_4) |G_c(t_3,t_4)|^{q/2-1} \, ,
\end{align}
and $dt_{1,2}$ and $dt_{1,2,3,4}$ are shorthands for $dt_1 dt_2$ and $dt_1 dt_2 dt_3 dt_4$ respectively. Note $\widetilde{K}_c$ is related to $K_c$, defined in \refb{kleb_frecursion}, by similarity transformation.

Next, we integrate $\sigma_\psi$ and $\sigma_\kappa$ out, to obtain
\begin{align}
\overline{Z} &= \int Dg_\psi Dg_\kappa~e^{-S_{eff}[g_\psi,g_\kappa]} \, ,
\end{align}
with
\begin{align}
\nonumber
S_{eff}[g_\psi, g_\kappa] &= \frac{NJ_0^2 (q-1)}{4\left[ 1- \epsilon(\frac{q}{p}-1)\right]} g_\psi \circ \left[ \widetilde{K}_c^{-1} -1 + \epsilon \frac{q-p}{q-1} \right] \circ g_\psi + \frac{n J_1^2 (q-1)}{4} g_\kappa \circ \left[ \widetilde{K}_c^{-1} - \frac{p-1}{q-1} \right] \circ g_\kappa \\
& - \frac{n J_0 J_1 (q-p)}{2 \left[ 1 - \epsilon (\frac{q}{p}-1)\right]^{1/2}} g_\psi \circ g_\kappa \, . \label{Seff}
\end{align}
Here $\circ$ denotes convolution product. E.g. 
\begin{align}
\nonumber
g_\psi \circ g_\psi &:= \int dt_1 dt_2 \, g_\psi(t_1,t_2) g_\psi(t_1,t_2), \\
g_\psi \circ \widetilde{K}_c^{-1} \circ g_\psi &:= \int dt_1 dt_2 dt_3 dt_4 \, g_\psi(t_1, t_2) \widetilde{K}_c^{-1} (t_1,t_2;t_3,t_4) g_\psi (t_3,t_4) \, .
\end{align}
It is straight forward to evaluate two point functions of $g_\psi$ and $g_\kappa$ from \refb{Seff}, which then leads to the four point functions $\mathcal{F}^\psi, \mathcal{F}^\kappa, \mathcal{F}^{\psi \kappa}$ using \refb{Gg}. 

To make use of the conformal symmetry let us define
\begin{align}
\nonumber
\mathcal{F}^\psi(t_1,t_2;t_3,t_4) &=: G_\psi^{saddle}(t_1,t_2) G_\psi^{saddle}(t_3,t_4) \mathcal{F}^\psi(\chi)\, , \\
\nonumber
\mathcal{F}^\kappa(t_1,t_2;t_3,t_4) &=: G_\kappa^{saddle}(t_1,t_2) G_\kappa^{saddle}(t_3,t_4) \mathcal{F}^\kappa(\chi) \, , \\
\nonumber
\mathcal{F}^{\psi \kappa}(t_1,t_2;t_3,t_4) &=: G_\psi^{saddle}(t_1,t_2) G_\kappa^{saddle}(t_3,t_4) \mathcal{F}^{\psi \kappa}(\chi) \, ,\\
-G_c(t_1,t_3) G_c(t_2,t_4) + G_c(t_1,t_4) G_c(t_2,t_3) &=: G_c(t_1,t_2) G_c(t_3,t_4) \mathcal{F}_{0,c}(\chi) \, . \label{ftofchi}
\end{align}
One gets to the following expressions:
\begin{align}
\mathcal{F}^\psi(\chi) &=  \frac{1}{  p -1- \epsilon (q-p) } \frac{ (q-1) - (p-1)\widetilde{K}_c }{(\widetilde{K}_c-1) \left( \widetilde{K}_c - \frac{q-1}{ p -1- \epsilon (q-p)}  \right)}  \mathcal{F}_{0,c} (\chi) \, , \label{fpsichi}\\
\mathcal{F}^\kappa(\chi) &= -\frac{ q -1 - \epsilon (q-p) }{ p-1 - \epsilon (q-p)} \frac{ \widetilde{K}_c - \frac{q-1}{q-1 - \epsilon (q-p)}  }{(\widetilde{K}_c-1) \left( \widetilde{K}_c - \frac{q-1}{p-1 - \epsilon (q-p)}\right)} \mathcal{F}_{0,c}(\chi) \, , \label{fkappachi}\\
\mathcal{F}^{\psi \kappa}(\chi) &=  - \frac{ (q-p) }{ p-1 - \epsilon (q-p)  } \frac{\widetilde{K}_c }{(\widetilde{K}_c-1) \left( \widetilde{K}_c - \frac{q-1}{p-1 - \epsilon (q-p)} \right)} \mathcal{F}_{0,c}(\chi) \label{fpsikappachi} \, .
\end{align}
Some comments are in order:
\begin{itemize}
\item
Appearance of $\widetilde{K}_c-1$ factor, in the denominators of \refb{fpsichi}, \refb{fkappachi}, \refb{fpsikappachi} is dictated by symmetry.
Spontaneous breaking of reparameterization symmetry implies vanishing action for Goldstones (more on this later). As in original SYK model, this necessitates the appearance of the $\widetilde{K}_c-1$ factor, in the denominators of \refb{fpsichi}, \refb{fkappachi}, \refb{fpsikappachi}, implying divergence of four point functions.
\item
Let us start with $\mathcal{F}^\psi$. Along with the original tower of primaries, resulting from $(\widetilde{K}_c-1)$ in the denominator, there is a new tower of primaries, coming from the new piece $\left( \widetilde{K}_c - \frac{q-1}{ p -1- \epsilon (q-p)}  \right)$ in the denominator. Appearance of these new primaries can be attributed to the interaction with probe fermions.
\item
$\mathcal{F}^\kappa$ and $\mathcal{F}^{\psi \kappa}$ have the same denominator as $\mathcal{F}^\psi$ and therefore lead to same spectrum of primaries. In this sense the
system can be said to have made a copy of the spectrum of primaries corresponding to the black hole. 
\item
$(\widetilde{K}_c-1)$ in the denominator implies all the four point functions have a non-conformal part. We shall revisit this later. 
\end{itemize}
For the sake of completeness, we give explicit forms of the conformal parts of these four point functions
\begin{align}
\nonumber
\mathcal{F}^\psi(\chi) &=  \frac{\alpha_0}{1+\epsilon} \sum_{m=1}^\infty \left[  \frac{h_m-1/2}{\pi \tan \pi h_m/2} \frac{ \Psi_{h_m}(\chi) }{k_c'(h_m)}  +  \epsilon \left(  \frac{q-1}{ p -1- \epsilon (q-p)} \right)^2 \frac{\widetilde{h}_m-1/2}{\pi \tan \pi \widetilde{h}_m/2} \frac{ \Psi_{\widetilde{h}_m}(\chi)}{k_c'(\widetilde{h}_m)} \right] \, , \\
\nonumber
\mathcal{F}^\kappa(\chi) &=  \frac{\alpha_0}{1+\epsilon} \sum_{m=1}^\infty \left[ \epsilon \frac{h_m-1/2}{\pi \tan \pi h_m/2} \frac{ \Psi_{h_m}(\chi) }{k_c'(h_m)}  +   \left(  \frac{q-1}{ p -1- \epsilon (q-p)} \right)^2 \frac{\widetilde{h}_m-1/2}{\pi \tan \pi \widetilde{h}_m/2} \frac{ \Psi_{\widetilde{h}_m}(\chi)}{k_c'(\widetilde{h}_m)} \right] \, , \\
\mathcal{F}^{\psi \kappa}(\chi) &= \frac{\alpha_0}{1+\epsilon} \sum_m \Bigg[ \frac{h_m-1/2}{\pi \tan \pi h_m/2} \frac{ \Psi_{h_m}(\chi) }{k_c'(h_m)} - \left( \frac{q-1}{ p -1- \epsilon (q-p)} \right)^2 \frac{\widetilde{h}_m-1/2}{\pi \tan \pi \widetilde{h}_m/2} \frac{ \Psi_{\widetilde{h}_m}(\chi)}{k_c'(\widetilde{h}_m)} \Bigg] \, , \label{fchi}
\end{align}
where $h_m$ is $m^{th}$ solution of $k_c(h)=1$ and $\widetilde{h}_m$ is $m^{th}$ solution of $k_c(h)=\frac{q-1}{p-1-\epsilon(q-p)}$, where $k_c(h)$ is given in \refb{kch}. For the detail of the function $\Psi_h(\chi)$, we refer the reader to \cite{Maldacena:2016hyu}. We take $h_0=2$ to be the non-conformal mode, which is excluded in the above expressions. Physics of this mode is related to the soft mode, which we shall embark upon presently.
%%%%%%%%%%%%%%%%%%
\subsection{Spectrum of primaries} \label{s3.3}
%%%%%%%%%%%%%%%%%%
In the limit $\chi \rightarrow 0$, four point functions \refb{fchi} reduce to
\begin{align}
\nonumber
\mathcal{F}^\psi(\chi) &\sim \sum_{m=1}^\infty \left[ c_m^2 \chi^{h_m} + \epsilon \widetilde{c}_m^2 \chi^{\widetilde{h}_m} \right] \, , \\
\nonumber
\mathcal{F}^\kappa(\chi) &\sim \sum_{m=1}^\infty \left[ \widetilde{c}_m^2 \chi^{\widetilde{h}_m} + \epsilon c_m^2 \chi^{h_m} \right] \, , \\
\mathcal{F}^{\psi \kappa}(\chi) &\sim \sum_{m=1}^\infty \left[ c_m^2 \chi^{h_m} - \widetilde{c}_m^2 \chi^{\widetilde{h}_m} \right] \, ,  \label{fchi0}
\end{align}
with 
\begin{align}
c_m^2 &=  \frac{\alpha_0 \left( h_m-1/2 \right)}{(1+ \epsilon) \pi k_c'(h_m) \tan \frac{\pi h_m}{2}}  ~,~
\widetilde{c}_m^2 =  \frac{\alpha_0 \left( \widetilde{h}_m-1/2 \right)}{(1+ \epsilon) \pi k_c'(\widetilde{h}_m) \tan \frac{\pi \widetilde{h}_m}{2}}  \, . 
\end{align}
This implies the existence of two towers of primaries with scaling dimensions $\{h_m\}$ and $\{ \widetilde{h}_m\}$. We call the corresponding primaries $\mathcal{O}_m$ and $\widetilde{\mathcal{O}}_m$. \refb{fchi0} can be reproduced from the operator product expansions (henceforth abbreviated as OPE)
\begin{align}
\nonumber
\frac{1}{N} \sum_i \psi_i(\tau_1) \psi_i(\tau_2) &= \frac{ G^{saddle}_\psi(\tau_{12})}{\sqrt{N}} \sum_m \left[  c_m |\tau_{12}|^{h_m} \mathcal{O}_m \left( \frac{\tau_1+\tau_2}{2}\right) - \sqrt{\epsilon} \widetilde{c}_m |\tau_{12}|^{\widetilde{h}_m} \widetilde{\mathcal{O}}_m \left( \frac{\tau_1+\tau_2}{2}\right) \right] \, , \\
\frac{1}{n} \sum_x \kappa_x(\tau_1) \kappa_x(\tau_2) &= \frac{ G^{saddle}_\kappa(\tau_{12}) }{\sqrt{n}} \sum_m \left[ \widetilde{c}_m |\tau_{12}|^{\widetilde{h}_m}  \widetilde{\mathcal{O}}_m \left( \frac{\tau_1+\tau_2}{2}\right) + \sqrt{\epsilon}  c_m |\tau_{12}|^{h_m}  \mathcal{O}_m \left( \frac{\tau_1+\tau_2}{2}\right) \right] \, . \label{OPE}
\end{align}
Note that the OPE of $\psi$ fields is dominated by $\mathcal{O}$ primaries, whereas the OPE of $\kappa$ fields is dominated by $\widetilde{\mathcal{O}}$ primaries,
%%%%%%%%%%%%%%%%%%%%%%%%%%%%%%%%%%%%%%%%%%%%%%%%%%%%%%%%%%%%%%%%%%%%%%%%%%%%%%%%%%
%%%%%%%%%%%%%%%%%%
\subsection{The soft mode} \label{ssoft}
%%%%%%%%%%%%%%%%%%
A Goldstone, referred to as the soft mode in the present context, is any variation of the fields, around the saddle, that still solves the Schwinger Dyson equations \refb{SDsaddlesigma}. The equations obeyed by such small variation is simply the Schwinger Dyson equation expanded around the saddle. For the bilocal local fields $G_\psi$ and $G_\kappa$, one has two equations to start with. But since both $G_\psi^{saddle}$ and $G_\kappa^{saddle}$ are proportional to $G_c$ and both self energies are proportional to $G_c^{q-1}$, they boil down to same equation 
\begin{align}
(K_c -1) \circ \delta G_c &= 0 \, . \label{Goldeqn}
\end{align}
This in particular implies that there are no new soft modes, due to the presence of probe fields.
Instead of solving \refb{Goldeqn} to get the soft modes, one can simply note that any reparameterization of the saddle (proportional to) $G_c$, is also a solution to saddle point equations. Thus small change of the saddle point field configurations under arbitrary reparameterization is a Goldstone. Such reparameterizations, with some rescaling, provide a basis for soft modes:
\begin{align}
g_m (\tau_1,\tau_2) &= \frac{i b^{q/2}}{q } \left( \frac{2\pi}{\beta}\right)^2 \frac{e^{-2i\pi my/\beta}}{\sin\frac{\pi x}{\beta}}  \left[ - m \cos \frac{2\pi m x}{\beta} +  \frac{\sin \frac{2\pi m x}{\beta}}{\tan \frac{\pi x}{\beta}} \right] \, . \label{Gn}
\end{align}
We have chosen to work in finite temperature and $x=\tau_1-\tau_2,~y = \frac{\tau_1 + \tau_2}{2}$. Note, $g_m=0$ for $m=0, \pm 1$. This is because these correspond to $SL(2, \mathbb{R})$ reparameterizations and hence are not Goldstones. It can be checked that
\begin{align}
g_m \circ g_n &= \delta_{m+n} \frac{b^q}{q^2} \left( \frac{2\pi}{\beta}\right)^4 \beta^2 \frac{|m|(m^2 -1)}{3} \, .
\end{align}
For a given infinitesimal reparameterization $\varepsilon(\tau) = \sum_m \varepsilon_m e^{-2\pi i m  \tau/ \beta}$, $g_\psi, g_\kappa$ have the following soft parts
\begin{align}
g_\psi(\varepsilon) &= \left[ 1 - \epsilon \left( \frac{q}{p} -1 \right)\right]^{1/2} \sum_m \varepsilon_m g_m \, , ~
g_\kappa(\varepsilon) = \frac{J_0}{J_1} \sum_m \varepsilon_m g_m \, .
\end{align}
All such $g_\psi(\varepsilon), g_\kappa(\varepsilon)$ are eigenfunctions of $K_c$ (equivalently $\widetilde{K}_c$ which is a similarity transformed version of $K_c$) with unit eigenvalue. This results in vanishing of the effective action \refb{Seff} in the soft sector. Consequently one gets divergent contributions from this sector to the two functions of $g_\psi, g_\kappa$-s, which in turn causes the divergence in four point functions of $\psi$ and $\kappa$-s.

A divergence in correlation functions is unacceptable in any physical theory. In the present case, the divergence has its origin in the vanishing action of the soft modes. Thus the only way out is to endow the soft modes with some small but non-zero action. However vanishing of the soft mode action is an artefact of spontaneous breaking of reparameterization symmetry, so a non-zero action would invariably imply the soft modes are not true Goldstones, but merely pseudo-Goldstones. Had the reparameterization symmetry been an explicit symmetry of the Hamiltonian, there would have been no way out. Luckily, the symmetry in hand is only an emergent one, obtained in a certain regime and if we go a little away from that regime, the deviation from the symmetry turns the Goldstones into pseudo-Goldstones. Here the relevant regime is deep infra-red, or equivalently large $J_0$. Thus moving away from this regime implies taking $1/J_0$ corrections into account. This was achieved in \cite{Maldacena:2016hyu}, where the following expression for the eigenvalues of $K_c$, corresponding to the pseudo-Goldstone $g_m$, was found
\begin{align}
k(2,m) &= 1 - \frac{\alpha_K |m|}{\beta \mathcal{J}} + \mathcal{O}(J_0^{-2})\, .
\end{align}  
Using this we find the soft mode action to be
\begin{align}
S^{soft} &= \frac{\alpha_K (N+n)}{12 \alpha_0 q^2 \beta \mathcal{J}} \sum_{|m| \geq 2} m^2 (m^2 -1) \varepsilon_m \varepsilon_{-m} 
=  \frac{\alpha_K (N+n)}{12 \alpha_0 q^2 \beta \mathcal{J}} \int_0^\beta d\tau \left[ (\varepsilon'')^2 - \left( \frac{2\pi}{\beta}\right)^2 (\varepsilon')^2 \right] \, . \label{softaction}
\end{align}
Notably this differs from the soft mode action of SYK model, by a multiplicative factor $(1+\epsilon)$. Since the functional form could not have been anything else by $SL(2,\mathbb{R})$ symmetry, there was really scope for only a multiplicative correction. What is far from obvious though is the absence of the coupling $J_1$ in this correction factor. 

We would like to mention that in the intermediate steps, terms of the form $\sum_m |m| (m^2-1) \varepsilon_m \varepsilon_{-m}$ appear but cancel at the end. Cancellation of such undesirable points to consistency of keeping the full $\epsilon$ dependence.

In the regime of interest, i.e. large $J_0$, \refb{softaction} is small, and vanishes as $J_0 \rightarrow \infty$. This small action translates into large, but finite, contributions to four point functions. We skip the details and jump to the dominant parts (denoted by the subscript ``big") of various four point functions:
\begin{align}
\nonumber
\frac{\mathcal{F}^\psi_{big}(\tau_1, \tau_2; \tau_3, \tau_4)}{G_\psi^{saddle}(\tau_1, \tau_2) G_\psi^{saddle}(\tau_3, \tau_4)} &= \frac{\mathcal{F}_{big}^\kappa(\tau_1, \tau_2; \tau_3, \tau_4)}{G_\kappa^{saddle}(\tau_1, \tau_2) G_\kappa^{saddle}(\tau_3, \tau_4)} = \frac{\mathcal{F}_{big}^{\psi \kappa}(\tau_1, \tau_2; \tau_3, \tau_4)}{G_\psi^{saddle}(\tau_1, \tau_2) G_\kappa^{saddle}(\tau_3, \tau_4)} \\ 
&= \left[ 1 + \epsilon \frac{q-p}{q-1}\right]^{-1} \frac{6 \alpha_0 \mathcal{J} \beta}{\pi^2 \alpha_K} \sum_{|m| \geq 2} \frac{e^{-2i\pi m (y_{12} - y_{34})/\beta}}{m^2 (m^2-1)} f_m (x_{12}) f_m(x_{34}) \, .\label{fbig}
\end{align}
Note that they are proportional to each other.
%%%%%%%%%%%%%%%%%%
\subsection{Chaos} \label{schaos}
%%%%%%%%%%%%%%%%%%
Finally we come the chaos. As in SYK model, the Lyapunov exponent can be determined by looking at the dominant part of the four point functions. There are three Lorentzian  four point functions to be looked at:
\begin{align*}
F^\psi(t) &:= \frac{1}{N^2} \sum_{j,k} \langle \psi_j(t + \frac{3i\beta}{4}) \psi_k(i\frac{\beta}{2}) \psi_j(t +i \frac{\beta}{4}) \psi_k(0) \rangle_L \, , \\ 
F^\kappa(t) &:= \frac{1}{n^2} \sum_{x,y} \langle \kappa_x(t +i \frac{3\beta}{4}) \kappa_y(i\frac{\beta}{2}) \kappa_x(t +i \frac{\beta}{4}) \kappa_y(0) \rangle_L \, , \\ 
F^{\psi \kappa}(t) &:= \frac{1}{nN} \sum_{j,x} \langle \psi_j(t +i \frac{3\beta}{4}) \kappa_x(i\frac{\beta}{2}) \psi_j(t +i \frac{\beta}{4}) \kappa_y(0) \rangle_L \, .
\end{align*}
These can be obtained by evaluating the following Euclidean four functions first,
\begin{align*}
F^\psi(\tau) &:= \frac{1}{N^2} \sum_{j,k} \langle \psi_j(\tau - \frac{3\beta}{4}) \psi_k(-\frac{\beta}{2}) \psi_j(\tau - \frac{\beta}{4}) \psi_k(0) \rangle_E \, , \\ 
F^\kappa(\tau) &:= \frac{1}{n^2} \sum_{x,y} \langle \kappa_x(\tau - \frac{3\beta}{4}) \kappa_y(-\frac{\beta}{2}) \kappa_x(\tau - \frac{\beta}{4}) \kappa_y(0) \rangle_E \, , \\ 
F^{\psi \kappa}(\tau) &:= \frac{1}{nN} \sum_{j,x} \langle \psi_j(\tau - \frac{3\beta}{4}) \kappa_x(-\frac{\beta}{2}) \psi_j(\tau - \frac{\beta}{4}) \kappa_y(0) \rangle_E \, .
\end{align*}
and then continuing to $\tau \rightarrow it$. Before we proceed, we recall that while evaluating Lyapunov exponent in SYK model previously, only the functional dependence of $F^\psi(\tau)$ on $\tau$, was of consequence. This allows us to make use of the fact that all three kinds of four point functions, and consequently $F^\psi(\tau), F^\kappa(\tau), F^{\psi \kappa}(\tau)$ are proportional to each other. Thus the Lyapunov exponent for any of them is same as that for $F^\psi(\tau)$. This has already been evaluated to be $2\pi/\beta$.
%%%%%%%%%%%%%%%%%%
\subsection{How small should a probe be?} \label{ssize}
%%%%%%%%%%%%%%%%%%
$H_{SYK}$ and $H_{probe}$ contribute to free energy at $\mathcal{O}(N)$ and $\mathcal{O}(n)$ respectively. Thus for $n<<N$, thermodynamics is dominated by SYK system and $\kappa$ fermions serve as probes. This may be called the probe regime, where general results of statistical mechanics are expected to hold, in particular SYK system is expected to work as heat bath for the probe system.

Now we draw the reader's attention to the fact that nowhere did we use the condition $n \ll N$, or equivalently $\epsilon \ll 1$, for the saddle point analysis. We implicitly assumed positivity of $1- \epsilon (\frac{q}{p}-1)$, i.e. $\epsilon < \frac{p}{q-p}$. However we can choose $\frac{p}{q-p}$ to be arbitrarily large, which in turn allows for arbitrarily large $\epsilon$. Therefore all the findings of this paper remain valid even if $n/N$ is not small, provided $q,p$ has been chosen accordingly. 

Let us consider the extreme case $n>>N$. Clearly as far as the thermodynamics is concerned, $\kappa$ system is expected to play the role of a heat bath and SYK system that of a probe. Spectrum of primaries are in accordance with this expectation. Dominant contribution to four point functions \refb{fchi0} in the OPE limit, comes from the $\widetilde{h}$ primaries and the OPE \refb{OPE} of fermion bilinears are dominated by corresponding primaries, i.e.  $\widetilde{\mathcal{O}}$-s. 

Curiously enough, this is not the case for chaos though. The $\kappa$ system, irrespective of its size, is a free system by itself and therefore not chaotic. It is intriguing then that an arbitrarily large $\kappa$ system is rendered maximally chaotic by arbitrarily small (in relative terms) SYK system. This has a moral  resemblance with Hawking radiation, in the sense black hole serves as the heat bath, despite being a finite dimensional system and the quantum field serves as probe, despite being infinite dimensional. although irrespective of any resemblance with Hawking radiation, it seems induction of chaos is worthy of study by itself, especially the curious possibility of induction of chaos by a smaller system onto a larger one.
%%%%%%%%%%%%%%%%%%%%%%%%%%%%%%%%%%%%%%%%%%%
\section{Summary and Future Directions} \label{scomment}
%%%%%%%%%%%%%%%%%%%%%%%%%%%%%%%%%%%%%%%%%%%
In this work we have presented a model for probes interacting with a near extremal black hole.
The black hole is modeled by $N$ number of Majorana fermions described by SYK model, whereas the probe comprises $n$ number of free Majorana fermions. The probe interacts with the SYK system through a SYK-like interaction. The full system is solvable in deep infrared and large $N$ as well as large $n$ limit. Interestingly, no assumption needs to be made about $n/N$. The probe, irrespective of its relative size, keeps the key features of the black hole intact, namely the emergence of reparameterization symmetry, pattern of its breaking and maximal Lyapunov exponent. However the OPE of SYK fermions now contains a new tower fo primaries, along with the old one. The new tower of primaries give subleading contribution when the probe is small. Curiously the OPE of the probe fermions turns out to contain the same set of fields. This can be interpreted as the probe copying some information of the black hole. The non-conformal part of the four point function can be traced back to the non-zero action for the pseudo-Goldstones, which continue to exist, since emergence and breaking of reparameterization symmetry is intact. The action for pseudo-Goldstones, or soft modes, is evaluated and found to be proportional to the corresponding action for SYK model, with the proportionality constant being independent of SYK-probe coupling. We lack a deeper understanding of this independence. The leading part of the non-conformal parts of the four point functions, when continued to out of time oder ones, diagnose chaos. All such four point functions are found to be maximally chaotic. This can be interpreted as the black hole rendering the probe maximally chaotic. 

It would be interesting to extend the present analysis to higher point function, to investigate to what extent does the probe encode relevant data about the black hole. Explicating aspects of thermalisation in this system would be another direction \cite{SS} worth exploring. Lastly, the finding that a small system can inflict chaos onto a much larger system, is stark contrast with the induction of temperature and calls for a better understanding. In particular it would be interesting to investigate if one can make any universal statement for infliction of chaos.

%%%%%%%%%%%%%%%%%%%%%%%%%%%%%%%%%%%%%%%%%%%%%%%%
\vspace{1cm} \noindent {\bf Acknowledgement:} This work was mostly supported by the CEFIPRA grant 5204-4 as well as J. C. Bose Fellowship of Rajesh Gopakumar, from the SERB, Govt. of India. Part of the work was supported by Laureate Award 15175 ``Modularity in Quantum Field Theory and Gravity" of the Irish Research Council. It is a pleasure to thank Bidisha Chakrabarty and Pinaki Banerjee for insightful discussions.
%%%%%%%%%%%%%%%%%%%%%%%%%
%%%%%%%%%%%%%%%%%%%%%%%%%
\appendix
%%%%%%%%%%%%%%%%%%%%%%%%%%%%%%%%%%%%%%%%%%%%%%%%%%
\section{Leading expressions of correlation functions from diagrammatic techniques} \label{app}
%%%%%%%%%%%%%%%%%%%%%%%%%%%%%%%%%%%%%%%%%%%%%%%%%%
For correlators involving only $\psi$ fields, i.e. $G^\psi$ and $\mathcal{F}^\psi$, leading diagrams are the same as in SYK model. Thus leading expressions for these correlators, obtained from diagrammatic techniques are also the expressions for these quantities obtained in SYK model. Therefore it suffices to check that our expressions for $G^\psi$ (same as $G_\psi^{saddle}$) and $\mathcal{F}^\psi$, given in \refb{saddlesol} and  \refb{fpsichi} reduce to their SYK counterparts in $\epsilon \rightarrow 0$ limit. This can be checked to be the case.

For correlators involving the probe fermions, a non-trivial check has to be performed. relevant correlators are given in equations \refb{saddlesol}, \refb{fkappachi}, \refb{fpsikappachi}. In the limit $\epsilon \rightarrow 0$, they reduce to  
\begin{align}
G_\kappa^{saddle} (t) |_{\epsilon=0} &= \left( \frac{J_0}{J_1}\right)^{2/p} G_c(t) \, , \label{gkapep0}\\
\mathcal{F}^\kappa(\chi) |_{\epsilon=0} &= 
%- \frac{q-1}{p-1} \frac{1}{\widetilde{K}_c - \frac{q-1}{p-1}} \mathcal{F}_{0,c}(\chi) = 
\frac{1}{1- \frac{p-1}{q-1} \widetilde{K}_c}  \mathcal{F}_{0,c}(\chi) \, , \label{fkapep0}\\
\mathcal{F}^{\psi \kappa}(\chi) |_{\epsilon=0} &= 
 %\frac{q-p}{p-1} \frac{\widetilde{K}_c}{(\widetilde{K}_c-1)(\widetilde{K}_c - \frac{q-1}{p-1})} \mathcal{F}_{0,c}(\chi) = 
 \frac{q-p}{q-1} \frac{\widetilde{K}_c}{(1- \widetilde{K}_c)( 1 - \frac{p-1}{q-1} \widetilde{K}_c)}  \mathcal{F}_{0,c}(\chi) \, . \label{fpsikapep0}
\end{align}
In the following, we shall reproduce these expressions from diagrammatic techniques.
%%%%%%%%%%%%%%%%%%%%%%%%%
%%%%%%%%%%%%%%%%%%%%%%%%%
\subsection{$G^\kappa$:} \label{appgkappa}
\begin{figure}[H]
	\begin{center}
		\includegraphics[scale=0.7]{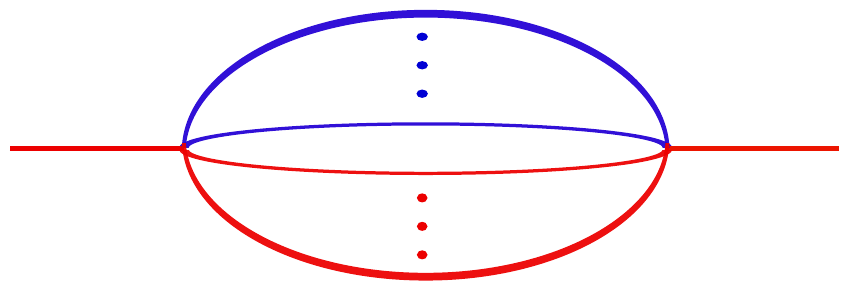}
		\caption{Simplest ``melonic" contribution to $\kappa$ propagator in large $N$ limit.}	\label{kappapropfig}
	\end{center}
\end{figure}

%%%%%%%%%%%
The simplest leading contribution to $G^\kappa : =\frac{1}{n} \sum_x \langle T \kappa_x(t) \kappa_x(0)\rangle$, the propagator for $\kappa$ fields, comes from figure \ref{kappapropfig}, which is obtained by cutting a $\kappa$ line in the right diagram of figure \ref{vacua}.
Since the right diagram of figure \ref{vacua} is $\mathcal{O}(n)$ and cutting a $\kappa$ line removes a factor of $n$, the diagram in figure \ref{kappapropfig} is $\mathcal{O}(1)$. Other leading contributions can be obtained by replacing $\kappa$ lines by figure \ref{kappapropfig} and $\psi$ lines by figure \ref{psiprop}. Following arguments similar to those used in deriving $G^\psi$, we have the following Schwinger-Dyson equation
\begin{align}
J_1^2 \int dt~G^\kappa (t_1,t) \left( G^\kappa(t,t_2) \right)^{p-1} \left( G^\psi(t,t_2) \right)^{q-p} = - \delta(t_1 -t_2) \, , \label{SDkappa}
\end{align}
which coincides with \refb{SDpsi} if we replace $J_1^2 \left( G^\kappa \right)^p$ by $J_0^2 \left( G^\psi \right)^p$. This implies the solution to \refb{SDkappa} is 
\begin{align}
G^\psi(t) = G_c(t) ~\text{and}~G^\kappa(t)= \left( J_0/J_1\right)^{2/p} G_c(t) \, , \label{kappaprop}
\end{align}
where $G_c$ is given in \refb{SYKpropa}. \refb{kappaprop} agrees with \refb{gkapep0}.

%%%%%%%%%%%%%%%%%%%%%%%%%
\subsection{$\mathcal{F}^\kappa$:} \label{appfkappa}
\begin{figure}[H]
	\begin{center}
		\includegraphics[scale=0.7]{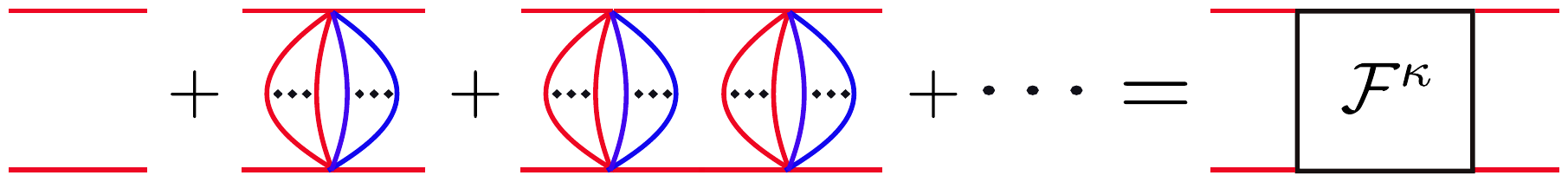}
		\caption{Simplest ``melonic" contribution to $\kappa$ four point function in large $N$ limit.}	\label{kappa4}
	\end{center}
\end{figure}

%%%%%%%%%%%%%%%%%%%%%%%%%
%%%%%%%%%%%
%\subsubsection{Leading contribution}
%%%%%%%%%%%
Gauge invariant four point function of $\kappa$ fields has the following structure:
\begin{align}
\frac{1}{n^2} \sum_{i,j=1}^n \left \langle \kappa_x(t_1) \kappa_x(t_2) \kappa_y(t_3) \kappa_y(t_4) \right \rangle &= G^\kappa(t_1,t_2) G^\kappa(t_3,t_4) + \frac{1}{n} \mathcal{F}^\kappa(t_1,t_2,t_3,t_4) \, . 
\end{align}
$\mathcal{F}^\kappa$ is given by the sum of the ladder diagrams in figure \ref{kappa4}.
Clearly, $\mathcal{F}^\kappa$ has same structure as in $\mathcal{F}^\psi$ and therefore
\begin{align}
\mathcal{F}^\kappa &= \frac{1}{1-K^\kappa} \mathcal{F}_0^\kappa \, ,
\end{align}
where
\begin{align*}
K^\kappa(t_1,t_2;t_3,t_4) 
&= -J_1^2 G^\kappa(t_1, t_3) G^\kappa(t_2, t_4) 
\left( G^\kappa(t_3,t_4)\right)^{p-2} (G^\psi(t_3,t_4))^{q-p} = \frac{p-1}{q-1} K^\psi (t_1,t_2;t_3,t_4) \, , 
\end{align*}
and $\mathcal{F}_0^\kappa = \left( \frac{J_0}{J_1} \right)^2 \mathcal{F}^\psi_0$. This gives
\begin{align}
\mathcal{F}^\kappa &= \left( \frac{J_0}{J_1}\right)^{4/p} \frac{1}{1- \frac{p-1}{q-1} K^\psi} \mathcal{F}^\psi_0 \, . \label{Fkap}
\end{align}
While deducing the expression for $\mathcal{F}^\kappa(\chi)$ from \refb{Fkap}, there are two factors of $G^\kappa$ which soak up $ \left( \frac{J_0}{J_1}\right)^{4/p} $ in the right hand side of \refb{Fkap} and thereby reproduce \refb{fkapep0}.
%%%%%%%%%%%%%%%%%%%%%%%%%
\subsection{$\mathcal{F}^{\psi \kappa}$:} \label{appfpsikappa}
\begin{figure}[H]
\begin{center}
\includegraphics[scale=0.7]{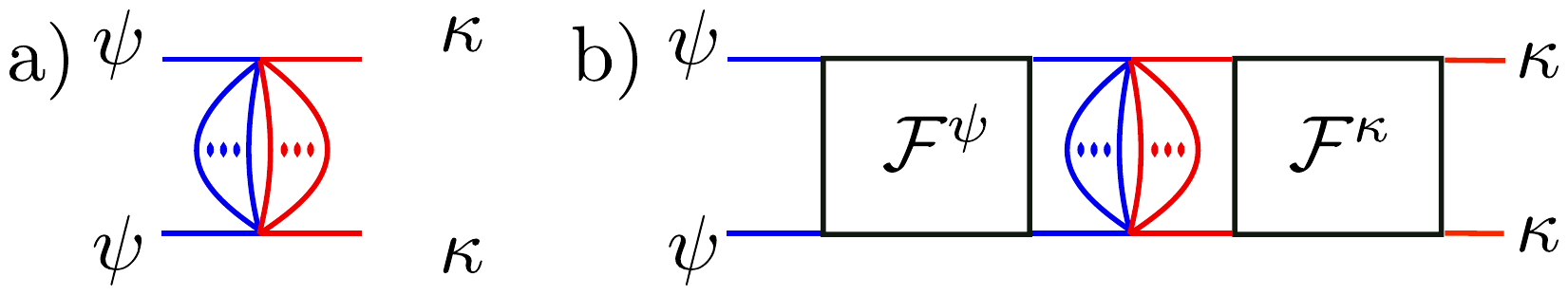}
\caption{Simplest ``melonic" contribution to mixed four point function in large $N$ limit.}	\label{psi2kappa2}
\end{center}
\end{figure}

%%%%%%%%%%%%%%%%%%%%%%%%%
%%%%%%%%%%%
%\subsubsection{Leading contribution}
%%%%%%%%%%%
Gauge invariant mixed four point function has the following structure 
\begin{align*}
\frac{1}{nN} \sum_{i=1}^N \sum_{x=1}^n \langle \psi_i(t_1) \psi_i(t_2) \kappa_x(t_3) \kappa_x(t_4) \rangle &= G^\psi (t_1,t_2) G^\kappa(t_3,t_4) + \frac{1}{N} \mathcal{F}^{\psi \kappa} (t_1,t_2;t_3,t_4) \, .
\end{align*}
Simplest leading contribution to $\mathcal{F}^{\psi \kappa}$ comes from the left diagram of figure \ref{psi2kappa2}.
Afterwards one can go on adding appropriate rungs in both sides of the ladder to obtain other leading diagrams. Summing these diagrams up, one has
\begin{align}
\mathcal{F}^{\psi \kappa} &= \left(\frac{J_0}{J_1}\right)^{2/p} \frac{(q-p)}{(q-1)} \frac{K^\psi}{(1-K^\psi) (1- \frac{p-1}{q-1}K^\psi)} \mathcal{F}_0^\psi \, . \label{Fpsikap}
\end{align}
While deducing the expression for $\mathcal{F}^{\psi \kappa}(\chi)$ from \refb{Fpsikap}, there appears a factor of $G^\kappa$, which soaks up $ \left( \frac{J_0}{J_1}\right)^{2/p} $ in the righthand side of \refb{Fpsikap} and thereby reproducing \refb{fpsikapep0}.
%%%%%%%%%%%%%%%%%%%%%%%%%%%%%%%%%%%%%%%%%%%%%%%%%%%%%%%%%%%%%%%%%%%%%%%%%%%%%%%%%%%%%%%
\providecommand{\href}[2]{#2}\begingroup\raggedright

\endgroup

\end{document}